\documentstyle[emulateapj,psfig]{article}

\accepted{December 5, 1997}
\journalid{337}{}
\articleid{}{}

\slugcomment{Accepeted to {\it The Astrophysical Journal (Letters)}}

\lefthead{Campusano, Kneib and Hardy}
\righthead{The arc-like object behind Abell cluster 3408}

\begin{document}
\hskip 3.5in{Version 1.2 \hskip 10pt 31 Dec 1997}
\title{
Gravitational lensing in low-redshift clusters of galaxies:\\
the arc-like object in Abell 3408 and its lensing interpretation}

\author {Luis E. Campusano$^1$}
\affil{Observatorio Astron\'omico Cerro Cal\'an, Departamento de Astronom\'{\i}a,
Universidad de Chile, Casilla 36-D, Santiago, Chile}

\author{Jean-Paul Kneib}
\affil{Observatoire Midi-Pyr\'en\'ees, 14 Av. E.Belin, 31400 Toulouse,
France}

\author{Eduardo Hardy}
\affil{D\'epartement de Physique, Observatoire du mont M\'egantic, Universit\'e Laval, Sainte-Foy, P.Q., Canada,
G1K 7P4}

\altaffiltext{1} {Visiting astronomer,
Cerro Tololo Inter-American Observatory, National Optical Astronomy Observatories,
which are operated by the Association of Universities for Research in
Astronomy, Inc. under contract with the National Science Foundation.}

\begin{abstract}
We analyze the seldomly discussed lensing effects expected in 
low-z clusters (z = 0.05-0.15), using as an example the  bright arc
($z=0.073$)  discovered by Campusano and Hardy(1996) near the central
elliptical galaxy of the cluster Abell 3408 ($z=0.042$). The
photometric and spectroscopic observations for both the central elliptical
and the arc are presented.
The mass distribution in A3408 is modeled 
by scaled versions of two representative
distributions derived from studies of clusters at higher
redshifts: 
i) a ``minimum'' mass case where the mass distribution follows
the light profile of the central elliptical galaxy and, ii) a
``maximum'' mass case where a typical massive 
dark halo is added to the previous case. The observed arc is
well reproduced by both models, but rather small
magnifications of the source galaxy are implied. The source galaxy
is tentatively identified in both the lensing and non-lensing scenarios as being a spiral. The smaller lensed spiral (14.6 ${h}_{50}^{-1}$ kpc,
$M_B$ = -18.2)
predicted by the dark halo model appears to fit 
the observations marginally better. Furthermore, we found that  
only the dark halo model predicts a measurable amount of
 weak shear in the images of faint background galaxies. We conclude
that observations, under very good seeing conditions,
of week shear in faint background galaxies in the direction of low-redshift
galaxy clusters are possible. The combination of the latter 
with X-ray data, can provide a powerful
tool to probe 
the mass distribution in the
very central region of galaxy clusters.

Subject headings: galaxies:clusters:general ---  gravitational lensing ---
galaxies:clusters: individual(Abell 3408)
\end{abstract}

\section{Introduction}

Clusters of galaxies are complex gravitationally bound systems
holding together three  components:  galaxies, ranging from the
faintest irregulars to the brightest elliptical galaxy that usually
sits at the center of the potential well,
hot gas [HG] (detected by its X-ray emission) and dark matter [DM].
Many recent studies of
the mass distribution of galaxy clusters exist, probing
the distribution with various tools:
galaxy dynamics (e.g. the CNOC survey, Carlberg et al 1997), X-ray analysis 
(e.g. Allen, Fabian \& Kneib 1996, Pislar et al 1997) and gravitational lensing
(e.g. Fort \& Mellier 1994, Narayan \& Bartelmann 1995).
Although, they are all generally consistent with a density profile that
goes like $\rho\sim1/r^2$ at intermediate scales (200-1000 kpc),
the exact profile at small and large scales is poorly known.
At small radii there is usually
a large number of galaxies in the central part of the clusters,
being difficult to disentagle the galaxy halo masses from 
the larger-scale cluster component. At large radii the data are
too noisy to make any safe statements. 
Numerical simulations by Navarro, Frenk \& White (1995, 1996) indicate
that simple gravitational systems follow a `Universal' mass profile
with a $\rho\sim1/r$ cusp at the center.  However this results is
subject to debate and other groups have claimed a steeper or shallower
central profile (Kravstov et al 1997, Moore et al 1997) .
Generally, due to the presence of the brightest central galaxy 
the total mass distribution will be cuspy (on a scale of a few kpc) 
since elliptical galaxies are well fitted by a 
de Vaucouleurs profile. Does the cusp exist only on the stellar
component, or is it also a property of the HG and DM components?
Solving these questions will help understand what the dynamical
interactions between the various components are, and whether any simple
scaling laws can be found.

The investigation of gravitational lensing of faint background galaxies
by low-redshift clusters of galaxies has been largely unexplored so far.
In fact, background galaxies  offer the possibility
of probing the mass distribution
of these clusters with higher-spatial resolution than that
allowed by high-$z$ cluster lenses, and should yield  
more accurate results on scales appropriate
to the study of the mass profile of the cluster core or even of individual galaxy
members.
The study of low-$z$ cluster lenses
has specially good prospect now that efficient wide-field cameras
will soon become available (e.g. the CFHT12K and the ACS aboard HST).

Theoretically, the cluster lensing
efficiency ---assuming a non-evolving mass profile with 
redshift, and a reasonable
redshift distribution for the faint galaxy population (see for example
Natarajan \& Kneib 1996) --- has its maximum around $z\sim 0.2$. Thus, a cluster
at $z\sim 0.1$ is as efficient, in terms of the
mean shear produced at a certain angular distance from its center,
as a $z\sim 0.3$ cluster. 
For low redshift clusters the lensing efficiency depends on the total 
mass profile at the center, whereas for high redshift clusters
it depends on the redhift distribution of the background faint galaxies.

There is already evidence of lensing in low redshift clusters of
galaxies, but
a discussion of the use of these cluster lenses for the investigation
of the mass profile of clusters has not yet been addressed. 
Allen, Fabian \& Kneib (1996) discovered a
$z=0.43$ arc in the PKS0745 $z=0.10$ cluster, which has been successfully
modelled as a gravitational lens image. 
A second case has been 
suggested by Shaya, Baum and Hammergren (1996), who found an
arc-like feature-- so far without redshift information -- close to 
NGC4881 ($z=0.024$) in the Coma cluster. And thirdly, there is
the tentative case found by  
Campusano \& Hardy (1996) of a bright $z=0.073$
arc-like feature in a $z=0.042$ cluster (Abell 3408). 
In this paper, we further investigate the lensing by low-$z$
clusters  by discussing the possibility that the arc in Abell 3408
is a gravitational image, and by computing the
elliptical distorsions expected in the images of the faint background
galaxies (weak shear) under different mass models. 
We adopt $\Omega_0$=1,
$\lambda$=0 and H$_0$= 50 $h_{50}^{-1} km~s^{-1}~Mpc^{-1}$; 
1" corresponds to 1.136 $h_{50}^{-1}$ kpc at the distance of the
cluster.

\newpage

\section{Observations}

Little is known about A3408 ($z=0.042$), whose J2000.0
coordinates are
RA = 07$^h$08$^m$29$^s$, 
DEC = -49$\arcdeg$12$\arcmin$50.3$\arcsec$. It lies in a largely
unexplored low galactic latitude section of the southern sky and it 
may even form a single unit with A3407 (Galli {\it et al.}
1993). Abell 3408 has been detected by the Rosat-All-Sky-Survey which
measured a flux $L_X = 0.5 \times 10^{44}$ erg~$s^{-1}$ (Ebeling et al 1996).  
Recently, Campusano and Hardy(1996) discovered  an
arc-like feature ($z=0.073$) near its center. We present here -- in full --
the photometric and spectroscopic observations of the arc and the central
elliptical galaxy of A3408.

\subsection{Photometry}

Imaging observations trough the B, R and I filters 
were obtained with the CF/CCD camera system at the
Cerro Tololo Inter-American Observatory (CTIO)
0.9-m telescope. The detector was the 2048 $\times$2048 Tek 2k \#3 CCD, which
has a scale of 0.40 $\arcsec$/pixel.
Landolt(1992) standard stars observation were used for the photometric
calibration. The discovery image was obtained serendipitously during observations of southern 
galaxy clusters as part of a project conducted in collaboration with Dale,
Giovanelli, \&  Haynes (Cornell).

Fig.~\ref{fig:BIfull} displays the combination of the B and I
images of the A3408 cluster. The field size is 6'$\times$8' and is
centered on the brightest elliptical of the cluster. 
An  inspection of Fig.~\ref{fig:BIfull} reveals
an arc-like object $\sim$ 50" away
from the brightest central elliptical galaxy; its orientation
suggesting the possibility of a gravitational lensing effect.
Fig.~\ref{fig:model} is a zooming of the R-band image
showing the arc and the central galaxy.
The width of arc-like object is unresolved, with a FWHM of 
$\sim$1.2"  compatible with the observed seeing. Its magnitude and
color are
(after removal of the contiguous stellar object from the images
via DAOPHOT) as follows : $R$ = 18.60, $R-I$ = 0.42, and $B-R$ = 0.79. The 
mean surface brightness values are ${\mu}_{B}$ = 24.23 mag/arcsec$^2$, 
${\mu}_{R}$ = 23.44 mag/arcsec$^2$, and ${\mu}_{I}$ = 23.02
mag/arcsec$^2$. The length of
the arc is $\sim$ 11".

Using the {\em ellipse} package in IRAF/STSDAS we computed the surface
photometry of the central elliptical in R and I band. The surface brightness
profile is well fitted by a
de Vaucouleurs profile with effective radius R$_e$=19.3$\pm 3$ arcsec
(22$\pm 4$ $h_{50}^{-1}$ kpc) and
$\mu_E(I)$=21.5 mag/arcsec$^2$, $\mu_E(R)$=22.3 mag/arcsec$^2$.
This galaxy has a small ellipticity $\varepsilon=(a^2-b^2)/(a^2+b^2)$=0.08
and an orientation of 35$\arcdeg \pm 5\arcdeg$. 
Its total B magnitude is 15.0 or M$_B$=-22.0 (L$_B$=1.0 $\times$ 10$^{11}$
L$_{B\odot}$).

\subsection{Spectroscopy}

Spectroscopic observations were carried out with the Ritchey
Chretien spectrograph at CTIO 4m telescope.
The detector was the Loral 3k CCD, which
after binning in the dispersion direction generated a readout format of
1595 pix$\times$785 pix.
We used a 350 lines mm$^{-1}$ grating blazed
at 4400 \AA\ giving a wavelength coverage of $\sim$3200-8800 \AA\ and a
spectral resolution of $\sim$8 \AA\ (1.5" slit). The position of the slit along the arc approached
the parallactic angle. Flux calibration was determined by observations
of the southern spectroscopic standards of Hamuy et al. (1994).

We obtained absorption spectra for the central elliptical  galaxy 
(yielding $z=0.0419\pm 0.0020$,
in good agreement with the results of Galli {\it et al.} 1993),
and for the stellar object inmediately next to the arc, which was found to
be a a K-type foreground star.

For the arc, its relatively high surface brightness yielded a high S/N
{\it integrated} spectrum (Fig.~\ref{fig:onedspec_arc}). 
Its derived mean redshift, based on the eight 
well-measured nebular emission lines (listed below), is 
$z=0.0728 \pm 0.0010$ corresponding to 21,840 $\pm 30$ km~ s$^{-1}$.
We also detected the absorption A-type spectrum of the underlying 
stellar population of the source which confirmed the emission-line
velocity from six absorption lines. Also, using the most prominent
nebular lines ( [OII]3727 \AA, [OIII]5007\AA\ and H$\alpha$) we
measured velocities along the arc, which revealed a velocity
gradient over a linear distance of about 10 kpc ${h}_{50}^{-1}$ with
$\sim 80~km~ s^{-1}$ peak-to-peak amplitude.

Equivalent widths (in \AA, indicated within parenthesis)
for the main nebular lines are as follows:
[OII] $\lambda$ 3727 (24.5); H$\beta$ (6.7); [OIII] $\lambda$ 4959 (8.7); 
[OIII] $\lambda$ 5007 (20.0); H$\alpha$ (58.7); [NII] $\lambda$ 6583 
(5.0); [SII] $\lambda$ 6717 (12.1). 
We are clearly not seeing a young starburst as the Balmer lines are not 
particularly broad -- H$\alpha$ and H$\beta$ both are
well fitted by gaussians with a FWHM of 8 \AA\ (i.e., 310 $ km~ s^{-1}$ ),
equivalent to the instrumental resolution-- , and the EW value of 6.7 \AA\ for H$\beta$ indicates an old age for the 
star-forming region from the models of Leitherer \& Heckman (1995). 
The overall emission spectrum is characteristic of intermediate-excitation 
HII regions with a ratio [OIII] $\lambda$ 5007/H$\beta$ = 3.0 , 
and low ratio [NII]/H$\alpha$ = 0.09. The underlying stellar spectrum with 
well-defined Balmer lines is compatible with an intermediate-A average 
spectral type. Thus, the spectral information is suggestive of a 
late-type spiral, an Irregular, or a blue compact dwarf
[BCD] galaxy.

The linear size of the arc 
is 20.6 ${h}_{50}^{-1}$ kpc, a value more compatible with 
a medium-size spiral than with a typical Irr or BCD,
but its rotational velocity is about half of that expected for an edge-on spiral
of this luminosity (Mathewson \& Ford. 1996). Notice also, that the
implied absolute magnitude of the arc-like object is
$M_B$ = -18.8, which is too faint for
a medium-size spiral but not incompatible with that of a small spiral such as M33, or
an irregular such as the LMC, both of which however have small linear sizes
of only a few(6-7) kpc. Also, the mean surface brightness  of the arc is too low for
a medium-size spiral, but again compatible with a small spiral,
an irregular, or a BCD.

Therefore, the straightforward interpretation of the arc implies a physical
object of somewhat contradictory properties. It appears to be an elongated
extragalactic object with emission characteristic of a typical 
HII region along most of its body, 
20.6 ${h}_{50}^{-1}$ kpc in size, a velocity profile with amplitude
$\sim$ 40 km~ s$^{-1}$, and a surface brightness and luminosity typical
of the much smaller (factor ~4) Magellanic irregular galaxies. 
A low luminosity can hardly be attributed to dust as the colors are normal. 
But, could  this arc-like object be the
product of gravitational lensing by the angularly close foreground
giant elliptical?

\newpage 

\section{Gravitational lensing in Abell 3408}

There is no evidence for a counter image of the arc-- although there are
objects of similar color to the arc, they are not at the position expected for
a ``two-image'' configuration. 
The ``gravitational radius'' of the arc is $\sim\
$ 50\arcsec\ (i.e. $\sim\ $ 56.8 ${h}_{50}^{-1}$ kpc at the cluster distance).

For a study of the gravitational lens scenario, we used both a simple 
spherical and a more
sophisticated elliptical lensing models in order to compute the projected mass of
the cluster within the radius defined by the bright arc.
All these potentials are able to produce a
single arc, as observed, but each one with different implications for
the source galaxy and the physical parameters of the cluster. 
We consider below two lensing mass model to compute the source properties
(see Table~\ref{tab:arcprop})
of the ``single'' arc:

i) {\it The constant M/L model}:
Given that the arc is at a distance of $\sim$3 R$_e$, we adopted
a mass distribution that
follows the light distribution -- assuming a constant M/L$_B$ $\sim$ 10 which is
consistent with the results from the fundamental plane --, and
which can be regarded as the
``minimum'' mass case. 
The mass enclosed within $R_{arc}$ is
M($R_{arc}$)$\sim$ 1.5 x 10$^{12}$ M$_\odot$, and the velocity dispersion of the
stars at R$_e$/2 is 225 km~$s^{-1}$.
The model predicts an amplification of 0.01 mag and a tangential
stretching of $\sim$1.02 -- there is obviously no much difference with respect to the no-lensing case.
The predicted weak-shear $\gamma$ at the distance of the arc, for
galaxies with $<z>\sim$0.7, is $\gamma\sim <\varepsilon_I>\sim $0.07
($<\epsilon_I>$ is the expected averaged mean ellipticity 
of the imaged faint galaxies - see {\it e.g.} Kneib et al 1996).
This value is at the detection limit.\\

ii) {\it The dark halo model}:
Both,
strong and weak lensing studies (e.g. Kneib et al 1996) clearly
indicate the presence of dark halos in clusters, centered on the brightest
elliptical galaxy, and with a typical M/L$_V$ ratio of $\sim$200.
Such dark haloes are also supported by the high velocity dispersion of
cluster galaxies, as well as by the increase of the velocity dispersion
of stars in giant elliptical galaxies (e.g. A2029, Dressler 1980). 
Adopting a massive mass profile comparable to the one found in A2218
(Kneib et al 1996) -- the ``maximum'' mass model --
we predict a single observed arc with an enclosed  mass 
of M($R_{arc}$)$\sim$ 4.4 x 10$^{13}$ M$_\odot$
corresponding to $M/L_B(R_{arc}$) $\sim$ 230, and an
asymptotic  velocity
dispersion for the member galaxies of 900 km~$s^{-1}$.
The model predicts an amplification of 0.6 mag and a tangential stretching 
of $\sim$1.4 for the source galaxy. 

In contrast to the prediction of the `minimum' mass model,
the expected weak-shear at the distance of the arc, for 
galaxies with $<z>\sim$0.7, is $<\varepsilon_I>\sim$0.5. This value
is quite large and readily measurable, thus it is a
strong prediction of this model. 

The velocity dispersion of the stars in the 
central elliptical galaxy is predicted to vary from
275 km~$s^{-1}$ at the center, for both the low-mass and dark-halo models,
to 225 km~$s^{-1}$ and 435 km~$s^{-1}$ at R$_e$/2, and
to 190 and 550 km~$s^{-1}$ at R$_e$=1,
for model i) and ii) respectively.

Note that this model should be considered as a fiducial one,
describing the mass distribution of A3408 only in an
approximate manner. In this exploratory work, the idea
was simply to extrapolate the mass distribution from a
higher-$z$  cluster and
scale it to a lower redshift.  The possibility of
adopting a more massive halo model, in order to get a better
agreement between the source galaxy a BCD, was considered
and discarded.
A much larger total mass was required, which would have placed 
this cluster amongst the most massive ones in the
Universe in spite of its relatively low X-ray luminosity.
 
\newpage

\section{Discussion and Conclusion}

But how common are the derived `source' galaxy characteristics? We are facing
two possibilities: a) an -- essentially ---
unlensed elongated galaxy of 20.6 ${h}_{50}^{-1}$ kpc
and $M_B$ = -18.8, or b) a lensed elongated source of
14.6  ${h}_{50}^{-1}$ kpc  and $M_B$ = -18.2.  We can derive a--priori
probabilities from an examination of the Binggeli, Sandage \& Tamman
(1988; Fig.~1) luminosity functions which also assume H$_0$= 50
~km~$s^{-1}~Mpc^{-1}$.

For case a) 35\% of the field galaxies will be Sc's with the required
absolute magnitude, but for the observed edge-on configuration only $\sim$
5-10\% are possible, and the probability drops significantly when the
low mass implied by the low rotation observed is considered. Even the
smallest Sc's have twice the observed rotational velocity.  For later
spiral types (Sd + Sm)the corresponding probabilities are about 50\%
lower. Irregulars or BCD's with this absolute magnitude are essentially
non-existent.  
For case b) the situation does not change significantly
due to the shallow slope of the observed LF: somewhat less than 30\% of
the field galaxies will be Sc's with the required magnitude, but for
the observed edge-on configuration only $\sim$ 5\% are possible; in
this case, the low mass implied by the low rotation observed is perhaps
a bit more realistic. At this absolute magnitude Irr and BCDs still
remain virtually non-existent. Both the mean surface ${\mu}_{B}$ = 24.23
mag/arcsec$^2$ -- although typical of the  Magellanic
irregular galaxies -- and color B-R = 0.79 of the arc are not
incompatible with those of small Sc galaxies. Then, a small lensed spiral seen
nearly face on, and thus of low projected rotational velocity, but
elongated by gravitational shear would fit the observations.

The above probabilities, however, are strictly {\it a-priori} and thus
must be taken with a grain of salt as we cannot at present exclude that
the observed arc is not simply a conspiring and unusual unlensed member
of the astronomical zoo.   We cannot at this point decide on the
identity of the source, but on statistical grounds a spiral seems
likely, although we cannot decide either on whether such a spiral would
or not be lensed, and if it is, what its intrinsic orientation would
be.

On a more positive note,
however, we have shown that there is a strong test that can be applied
to  the dark halo model:  deep wide-field imaging of low-$z$
clusters (similar to A3408) should display a weak shear easily
measurable with standard techniques. The analysis of such weak shear,
coupled with X-ray observations of these clusters, should unveil the
profile of the different mass components in the very central region of
the cluster of galaxies. A better understanding of the dynamics of these relaxed
systems would then be possible.

\acknowledgements
This research was supported by ECOS-CONICYT grant C95U02, by FONDECYT
grant No. 1970735 to LEC, by CNRS to JPK, and by NSERC-Canada and 
FCAR-Qu\'ebec to EH. We acknowledge the excellent 
technical support of the CTIO staff and 
thank Guy Mathez for his encouragement.

\newpage
 
\begin{table*}
\begin{tabular}{llllll}
\noalign{\smallskip}
\tableline
\noalign{\smallskip}
Model &  R$_{intr}$ & $\Delta m$ & Length & M$_B$ & Comments \\
      &    (mag)     &  (mag)      & (arcsec/kpc) & (mag)  &  \\
\noalign{\smallskip}
\tableline
\noalign{\smallskip}
Unlensed  & 18.6 & 0 & 11.0 / 20.6 & -18.8 &  \\
Constant M/L & 18.6 & 0.01 & 10.8 / 20.3 & -18.8 & M/L$_B=10$ and
$<\varepsilon_I>\sim$0.07. \\
Dark halo & 19.2 & 0.60 & 7.8 / 14.6  & -18.2 & M/L$_B\sim$ 230 and 
$<\varepsilon_I>\sim$0.50.
 \\
\noalign{\smallskip}
\tableline
\noalign{\smallskip}
\end{tabular}
\caption{
The implied intrinsic properties of the arc under three hypotheses:
no-lensing and lensing under two mass profiles. 
The predicted mean ellipticity of the faint background galaxies 
($<\varepsilon_I>$)
and the Mass/Luminosity ratio, both evaluated at the distance of the
arc from the giant elliptical galaxy, are given for the lensing models.
}
\label{tab:arcprop}
\end{table*}

\newpage

\begin{figure}[h]
\caption{6'$\times$8' field showing the central region of the A3408
cluster. An arc-like object is seen $\sim$50" away from the brightest
central elliptical galaxy. The pseudo-colors, obtained from  B and I CCD
frames taken with the 0.9-m CTIO telescope, were adjusted so that they
approximately match the real color contrast of the objects. North is at
the left and East at the bottom.
}
\label{fig:BIfull}
\end{figure}

\newpage

\vspace*{1cm}
\begin{figure}[h]
\psfig{file=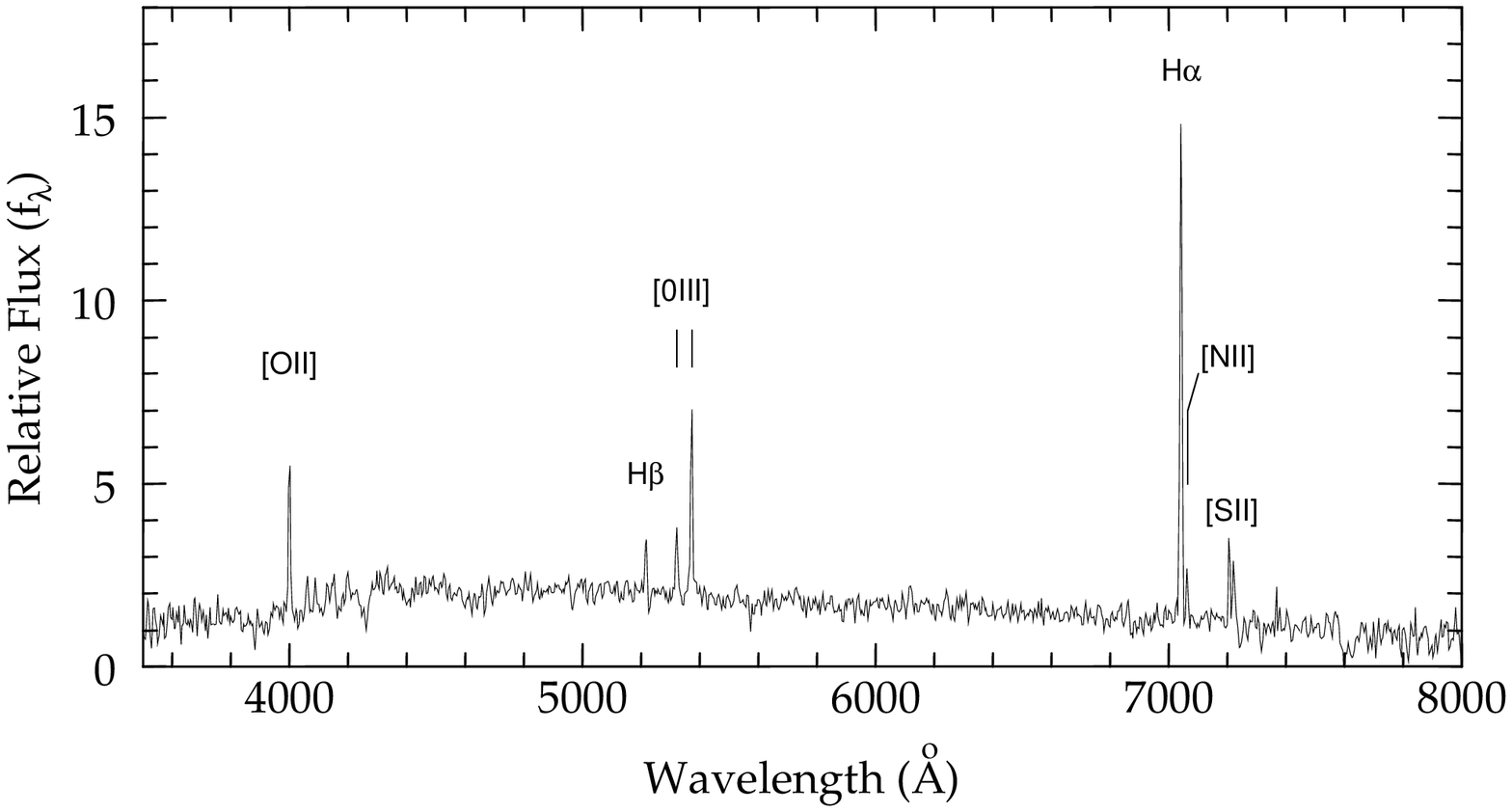,angle=270,width=\textwidth}
\caption{Optical spectrum of the arc-like object. A redshift 
of $z=0.0728\pm0.001$ is derived
from eight well measured nebular emission lines (marked).}
\label{fig:onedspec_arc}
\end{figure}
\newpage

\begin{figure}[h]
\psfig{file=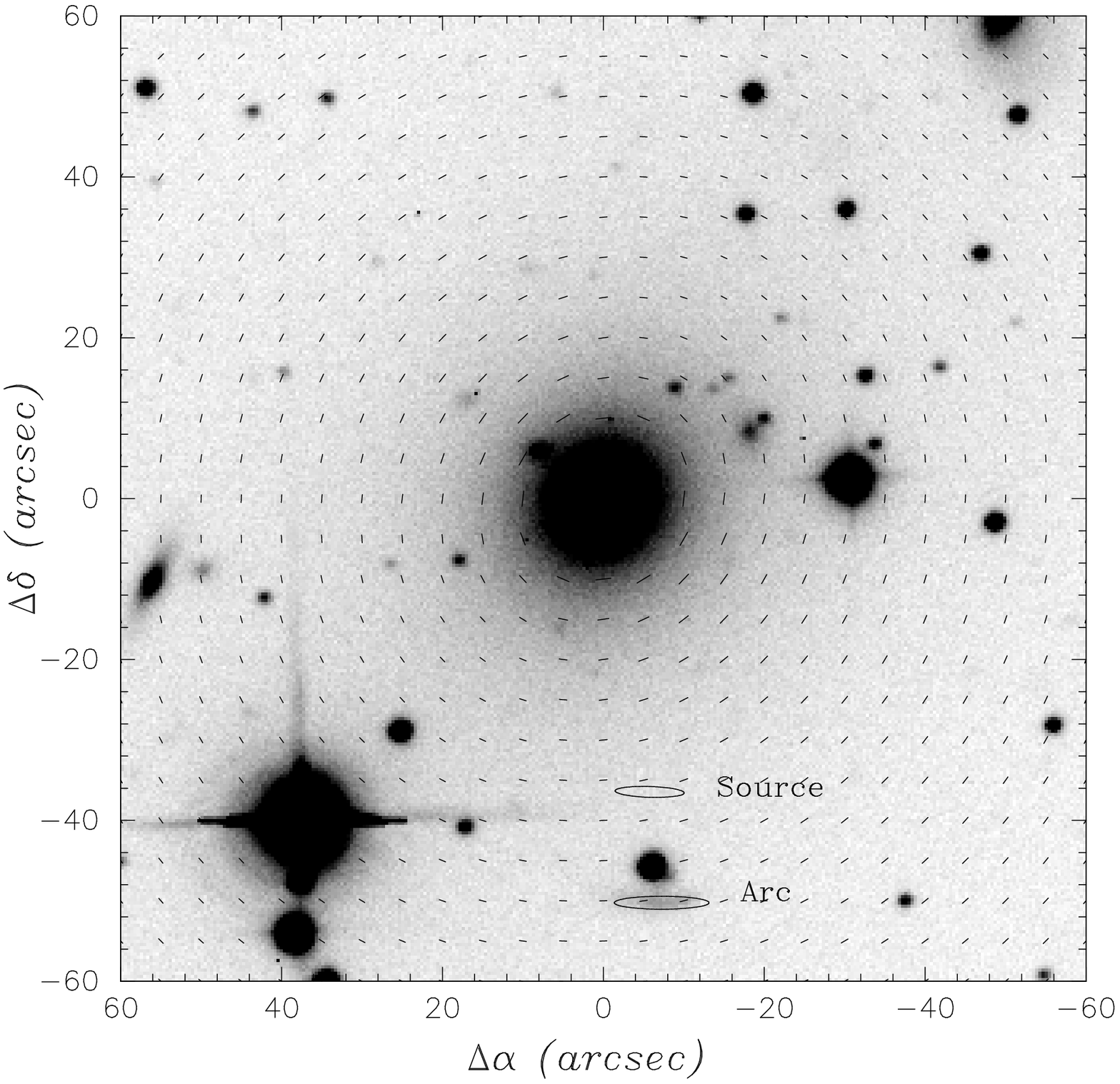,width=\textwidth}
\caption{Dark halo model. Zooming of R-band image taken with the
0.9-m CTIO telescope. The observed arc is at the
bottom, while the derived position of the source is marked with an ellipse.
The predicted shear field, for background galaxies with a mean redshift
of $\sim 0.7$, is overlaid; it ranges from  $<\varepsilon_I>\sim 0.3$
in the corners to  $<\varepsilon_I>\sim 0.7$ in the central part.
North is at the top and East at the left.
}
\label{fig:model}
\end{figure}

\newpage

\end{document}